\documentclass[pra,aps,reprint,showpacs,flushbottom]{revtex4-1}
\usepackage{xspace,amsmath,amsfonts,amsthm,amssymb,amsbsy,graphicx,color}

\begin{document}

\newcommand{\nn}{\nonumber}
\newcommand{\ms}[1]{\mbox{\scriptsize #1}}
\newcommand{\dg}{^\dagger}
\newcommand{\smallfrac}[2]{\mbox{$\frac{#1}{#2}$}}
\newcommand{\la}{\langle}
\newcommand{\ra}{\rangle}
\newcommand{\ket}[1]{| {#1} \ra}
\newcommand{\bra}[1]{\la {#1} |}
\newcommand{\pfpx}[2]{\frac{\partial #1}{\partial #2}}
\newcommand{\dfdx}[2]{\frac{d #1}{d #2}}
\newcommand{\ioh}{-\frac{i}{\hbar}}
\newcommand{\ohh}{-\frac{1}{\hbar^2}}
\newcommand{\half}{\smallfrac{1}{2}}

\title{Typical, finite baths as a means of exact simulation of open quantum systems}

\author{Luciano Silvestri$^1$}
\author{Kurt Jacobs$^2$} 
\author{Vanja Dunjko$^2$}
\author{Maxim Olshanii$^2$} 

\affiliation{$^1$Department of Physics, Boston College, Chestnut Hill, MA 02467, USA \\
$^2$Department of Physics, University of Massachusetts at Boston,
Boston, MA 02125, USA}

\begin{abstract} 
There is presently considerable interest in accurately simulating the evolution of open systems for which Markovian master equations fail. Examples are systems that are time-dependent and/or strongly damped. A number of elegant methods have now been devised to do this, but all use a bath consisting of a continuum of harmonic oscillators. While this bath is clearly appropriate for, e.g., systems coupled to the EM field, it is not so clear that it is a good model for generic many-body systems. Here we explore a different approach to exactly simulating open-systems: using a finite bath chosen to have certain key properties of thermalizing many-body systems. To explore the numerical resources required by this method to approximate an open system coupled to an infinite bath, we simulate a weakly damped system and compare to the evolution given by the relevant Markovian master equation. We obtain the Markovian evolution with reasonable accuracy by using an additional averaging procedure, and elucidate how the typicality of the bath generates the correct thermal steady-state via the process of ``eigenstate thermalization''. 
\end{abstract}


\maketitle 

\section{Introduction}

Many important applications of quantum mechanics involve small systems that are coupled to a large environment that acts as a thermal bath. The challenge of simulating the behavior of these systems is the enormity of the environment. If the damping induced by the bath is sufficiently weak, and the Hamiltonian of the system is constant, then simple Markovian master equations can be derived for the system density matrix alone~\cite{Breuer07}. While these are useful for weakly-damped systems, there are many situations in which they are insufficient. Even when more complex master equations can be derived outside of the weak-coupling regime~\cite{Breuer06, Albash12}, their accuracy is often difficult to determine without exact simulations~\cite{Breuer06, Fleming11, Rivas10}. Applications of current interest that require simulation of open systems beyond that of simple Lindblad equations include the coherent dynamics of photo-synthetic complexes~\cite{Mohseni08, Thorwart09, Caruso09, Sarovar11}, and coupled qubits under time-dependent control~\cite{Albash12}.  

In the last few years a number of numerical methods have been devised to simulate, essentially exactly, the dynamics of open systems coupled to infinite environments. This is possible because the infinite environment can be very well approximated by a system with fewer degrees of freedom, and in such a way that the accuracy of the approximation can be checked. These methods include the hierarchy of coupled master equations developed by Ishizaki and Tanimura~\cite{Ishizaki05}, those of Bulla~\textit{et al.\ }\cite{Bulla03} and the group of Plenio~\cite{Prior10, Chin10} that use renormalization-group techniques, and the path-integral method of Makri and Makarov~\cite{Makri95b, Makri95c}. All these methods provide essentially exact simulations of a system coupled to a specific kind of bath, that of a continuum of harmonic oscillators. This particular bath has become the standard for modeling open quantum systems, essentially by default. While it certainly applies to an atom coupled to the modes of the electromagnetic field, it is not so clear that it correctly models an open system strongly coupled to some ``generic'' many-body system. The assumption, of course, regardless of what bath model one is using, is that there is such a thing as a ``generic'' bath. For this to be true there must be a large class of many-body systems that produce the same behavior in small systems to which they are coupled. While this is true for weak coupling due to Fermi's golden rule, the question is open for strong coupling.  

Here we explore the possibility of exactly simulating an open system by using a very different kind of bath.  The bath we use, which we will refer to a ``typical'' thermal bath, is designed to possess certain key properties of many-body thermal baths. To the extent that baths consisting of thermalizing many-body systems induce a universal behavior in small systems to which they are coupled, one expects our bath to reproduce this behavior. The question of the universal behavior of thermal baths for strong coupling could be explored by comparing simulations of the standard harmonic oscillator bath with the bath we consider here. To determine the numerical resources required to simulate an open system with our bath, we compare our exact simulations for a weakly-coupled system with the evolution given by the Markovian Redfield equation for the same system~\cite{Redfield57, Redfield65, Breuer07}. 

Our choice of bath draws from an understanding of the structure of thermalizing many-body systems that emerged initially with the work of Srednicki~\cite{Srednicki94, Srednicki99} and Deutsch~\cite{Deutsch91}, and is related to Berry's conjecture~\cite{Srednicki94, Berry77}. The essential observation is that (almost) all of the eigenstates of a (thermalizing) large system reproduce the properties of the microcanonical ensemble at their respective eigen-energies (this microcanonical ensemble is the completely mixed state within a narrow energy band about the given eigen-energy), and thus places every small system in a canonical equilibrium state. It was noted also by those studying chaos that (almost all) the eigenstates of random Hamiltonians have the same property, which they described as ergodicity, and that random Hamiltonians therefore reproduce thermal (ergodic) behavior~\cite{Feingold86, Horoi95, Zelevinsky96, Flambaum96, Flambaum97b, Jacquod97, Tasaki98}. The ergodicity of the eigenstates of certain (non-random) multi-body Hamiltonians was also investigated in the context of understanding thermalization~\cite{Flambaum94, Horoi95, Borgonovi98}. Much more recently it was shown by Popescu, Short and Winter~\cite{Popescu06}, and Goldstein \text{et al.}~\cite{Goldstein06} (see also~\cite{Dong07, Goldstein10a, Goldstein10b, Reimann12, Short12}), that almost every pure state within a narrow energy band will behave as the microcanonical ensemble, from which it follows that random states will have the same property. Since almost all states have this ergodic property, states that do are called \textit{typical} states. It has been conjectured that all many-body systems that thermalize have eigenstates that are almost all typical. This conjecture is called the ``eigenstate thermalization hypothesis'' (ETH), a term coined by Srednicki~\cite{Srednicki94}, and it is supported by all numerical studies that have been performed to-date~\cite{Rigol08, Santos10a, Santos10b, Santos12, Genway12, Zhang12, Dubey12}. 

We choose the bath so that the combined system (by convention ``the universe''), consisting of the small system and the bath, has typical states. This is achieved by choosing the bath operator that couples to the system to be a random matrix, and we explain below why this generates eigenstate thermalization for the universe. The notion that a bath with typical states, also referred to as a ``random-matrix bath''~\cite{Gemmer06}, will provide a good model of a thermal bath is not new. There have been a number of studies showing that a random-matrix bath will induce damping and thermalization, and approximate master equations have been derived from these baths~\cite{Massimiliano03, Massimiliano03b, Lebowitz04, Gemmer06, Breuer06, Bartsch08} (see also~\cite{Rossini06} which uses a ``chaotic'' bath). Breuer, Gemmer, and Michel used a typical bath to obtain a simulation of a single qubit interacting with a thermal environment~\cite{Breuer06}. What has not been attempted before is to construct a bath to accurately simulate an arbitrary open system coupled to a thermal environment. As per the fundamental assumptions of statistical mechanics, this requires that the density of energy eigenvalues of the bath increases exponentially with energy~\cite{Riera12}. The fact that the bath energy levels must be sufficiently dense with respect to those of the system, and that the bath must also have a total energy range that is at least twice that of the system (see below) places lower limits on the size of the bath. 

In Section~\ref{model} we present the details of the bath model, and in Section~\ref{eigtherm} explain why it can be expected to thermalize the system via the mechanism of eigenstate thermalization. In Section~\ref{num} we present numerical results demonstrating the resulting thermalization for an arbitrary 4-level system and a bath of 5000 states. In Section~\ref{ent} we discuss the increase in the thermodynamic entropy of the system and bath during the equilibration, and how this is related to the entanglement between them. In Section~\ref{relax} we show that by averaging over many initial states of the bath, our simulation reproduces the relaxation given by the standard weak-coupling rate equations, namely the Markovian Redfield master equation~\cite{Redfield57, Redfield65, Breuer07}. We also discuss the question of when any bath model, and especially random-matrix models, might reproduce potentially universal relaxation induced by real many-body systems. Section~\ref{sumup} concludes with a summary of our results.

\section{The Model}
\label{model}

Our model consists of a small system (from now on ``the system'') coupled to a large system that we call the bath. The combined system is the tensor product of the system and bath, and we will refer to it as the universe. The Hamiltonian of our universe is given by 
\begin{equation}
    H_{\mbox{\scriptsize unv}} = H_{\mbox{\scriptsize sys}} + \hbar g X_{\mbox{\scriptsize sys}} \otimes Y_{\mbox{\scriptsize bath}} + H_{\mbox{\scriptsize bath}} , 
\end{equation}
where $H_{\mbox{\scriptsize sys}}$ is the system Hamiltonian, $H_{\mbox{\scriptsize bath}}$ is the bath Hamiltonian, $X_{\mbox{\scriptsize sys}}$ is the system coupling operator, $Y_{\mbox{\scriptsize bath}}$ is the bath coupling operator, and $g$ is a constant setting the overall size of the coupling. In what follows we always work in the joint energy-eigenbasis of the system and bath, so that $H_{\mbox{\scriptsize sys}}$ and $H_{\mbox{\scriptsize bath}}$ are diagonal. We also need to distinguish between the energy eigenstates of the universe when the interaction is turned off (these are merely the tensor-products of the energy eigenstates of the system and the bath), and the energy eigenstates of the universe when the interaction is on. We will refer to the former as the universe ``basis states'', and and those with the interaction turned on as the universe energy eigenstates. 

Since the bath must thermalize any system, it is the properties of the bath, along with those of $Y_{\mbox{\scriptsize bath}}$, that are the key to obtaining thermal behavior. The properties of our bath are as follows:  

1) The density of states of the bath: the bath must be chosen to have a density of energy eigenstates that increases exponentially with energy. This condition is essentially just the usual equilibrium thermodynamic assumption: the Boltzmann distribution for a small system in contact with a bath results directly from the assumptions that 1) the density of states of the bath is exponential as a function of energy, 2) that the energy of the universe is conserved, and 3) that all states of the universe are equally likely. The temperature of the bath is given by $T = 1/(k_{\mbox{\scriptsize B}} \beta)$, where the energy-density of states is $D(E) \propto \exp(\beta E)$. By definition, the temperature of a thermal bath should not change as energy is added (the bath is ``big''), which means merely that $\beta$ is a constant, independent of $E$. 

\textit{Note:} In fact, many-body systems have a density of states that peaks in the middle of the spectrum. (Consider for example a collection of spin-half particles in a magnetic field: at the maximum and minimum energy the particles are either all up or all down, so that there is only one state. Conversely there are many states in which exactly half the particles are up, and thus when the energy in the middle of the spectrum.) The reason that many-particle systems obey thermodynamics is that in practice, unless specially prepared, the states of these systems are always in the lower half of the spectrum where the density of states increases exponentially with energy. 

\begin{figure}[t] 
\leavevmode\includegraphics[width=1\hsize]{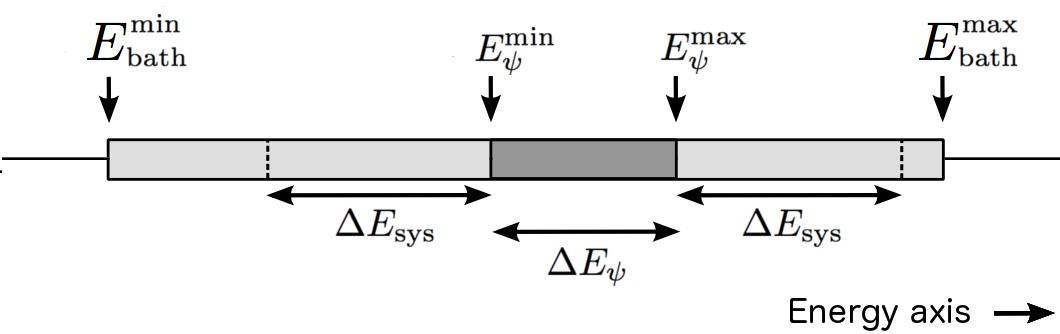} 
\caption{Here we show the relationships between the energy ranges of the bath, the system, and the initial state of the bath. The energy range of the bath is the interval $[E_{\ms{bath}}^{\ms{min}}, E_{\ms{bath}}^{\ms{max}}]$, depicted by the light gray region in diagram. The initial state of the bath is chosen to be a random superposition of all the bath energy eigenstates with energies in the interval $[E_{\psi}^{\ms{min}}, E_{\psi}^{\ms{max}}]$, for a total energy width of $\Delta E_\psi = E_{\psi}^{\ms{max}} - E_{\psi}^{\ms{min}}$. The energy range of the initial state is denoted byt he dark grey region in the diagram. The energy range of the system is denoted in the diagram by $\Delta E_{\ms{sys}}$. The energy windows given by the light grey areas on either side of the dark grey region must be at least as wide as the energy range of the system. This is so that the system has the ability to dump all its energy into the bath, and extract all its energy from the bath. Without this ability the bath cannot thermalize the system to the Boltzmann state.} 
\label{fig1} 
\end{figure}

2) The energy range of the bath: this must be large enough that the system can explore all its state-space while conserving the energy of the universe. Thus the system must be able to dump all its energy into the bath, and conversely absorb the same amount of energy from the bath. In choosing our system (below) we make an essentially arbitrary choice for the total energy range of the system, which is $\Delta E_{\ms{sys}} = 3.5 \hbar \mu$. Here $\mu$ sets the overall energy scale of the simulation. We will also have to choose the initial state of the bath so that it overlaps with a relatively large number of the bath energy eigenstates. If our bath was infinitely large we would not have to do this; a single initial energy state would suffice. But because our bath is not especially large, choosing the initial state to overlap with many bath states allows more averaging in the dynamics,  reducing the random fluctuations in the evolution. Let us say that our initial bath state overlaps with all the bath energy eigenstates with energies in the interval $[E_{\psi}^{\ms{min}}, E_{\psi}^{\ms{max}}]$. To ensure that the system can dump energy $\Delta E_{\ms{sys}}$ into the bath, given this initial state, the maximum bath energy $E_{\ms{bath}}^{\ms{max}}$, must be no less than $E_{\psi}^{\ms{max}} + \Delta E_{\ms{sys}}$. Similarly, the minimum bath energy must be no greater than $E_{\psi}^{\ms{min}} - \Delta E_{\ms{sys}}$. This relationship between the various energy ranges is depicted in Fig.~\ref{fig1}. 

We must also ensure that the energy states of the bath that play a role in the evolution are sufficiently densely packed in energy. The reason is that in order to thermalize the system, each energy eigenstate of the universe must contain a reasonable number of adjacent universe basis states. In particular, each universe energy eigenstate must be a typical state within a narrow energy window of the universe. By ``narrow'' we mean that the window is smaller than the energy gaps between the states of the system. The universe eigenstates will only overlap with a large number of universe basis states if the interaction, being on the order of $\hbar g$, is strong enough to mix many adjacent basis states. Thus the energy gaps between adjacent basis states must be much less than $\hbar g$, and for weak coupling $\hbar g$ must be much less than the gaps between the system states. Thus the energy levels of the bath must be dense compared to those of the system. We chose the interaction rate $g = 5\times 10^{-3}$.

We choose a bath of $5000$ states, and set the lowest energy to be $3 \hbar \mu$. To obtain a spectrum whose density increases exponentially, we start at the lowest level, and add levels one at a time. If the last energy level added has energy $E$, then the next energy level is chosen to have energy $E + \hbar \mu e^{-\beta E}$. Starting with $E = 3 \hbar \mu$, and adding 5000 levels, the maximum energy level is $E_{\ms{bath}}^{\ms{max}} \approx \hbar \mu 20$. With these choices the lowest 100 or so levels are not very dense, so we chose the energy range of the initial bath state to be $\hbar\mu[12.4,14.1]$. This means that the lowest energy of the bath explored during the evolution will be approximately $\hbar \mu (12.4 - 3.5) = \hbar \mu 8.9$, and the highest bath energy will be approximately $\hbar\mu (14.1 + 3.5) = \hbar \mu 17.6$. 

It is important to note that for larger values of $\beta$ (lower temperatures), the density of the bath energy states will be more skewed. This means that for a given size of the bath, and a given energy range for the bath, the lower energy levels will become more sparse with decreasing temperature. For lower temperatures we will therefore have to use larger baths, and so the numerical resources will increase as the temperature decreases. We discuss this further in the next section. 

3) The initial state of the bath: for the reasons explained in 2), we choose the initial state of the bath to contain the $350$ contiguous energy eigenstates that span the interval $\hbar\mu[12.4,14.1] $. Since our purpose in having the initial state contain many bath eigenstates is to reduce fluctuations via the resulting averaging, we choose the initial state to be a random superposition of these eigenstates. In particular, we choose all the coefficients in the superposition to have equal amplitudes and independently chosen random phases. 

4) The bath interaction operator: This operator, which we denote by $Y_{\mbox{\scriptsize bath}}$,  requires some complexity --- that is, its elements should, at least locally, vary in a more-or-less random fashion. So long as the elements of the interaction operator, $g X_{\mbox{\scriptsize sys}} \otimes Y_{\mbox{\scriptsize bath}}$, are large enough to mix together a significant number of adjacent universe basis states, this randomness ensures that the eigenstates of the universe are typical states within narrow energy bands. This typicality ensures in turn that the bath will thermalize the system (see Section~\ref{eigtherm} below). The random nature of the interaction operator is simple to achieve by choosing the off-diagonal elements of $Y_{\mbox{\scriptsize bath}}$ to be Gaussian random numbers with unit variance. We set the diagonal elements of $Y_{\mbox{\scriptsize bath}}$ to zero, so as to minimize their effect on the bath spectrum. (When the diagonal elements are zero, $Y_{\mbox{\scriptsize bath}}$ modifies the bath spectrum only to second order in perturbation theory, rather than first order). We also choose the interaction, along with all other contributions to the universe Hamiltonian, to be real. This reduces the numerical overhead in diagonalizing the Hamiltonian.  



The average size of the elements of $Y_{\mbox{\scriptsize bath}}$ should be uniform, in order to reproduce, in the weak-damping (Markovian) limit, the result that the relaxation rates are independent of the initial state of the system~\cite{Gemmer06}. Further, this is implied by the structure of many-body baths: in this case the system interacts only with its nearest neighbors, and even though there may be many of these it cannot immediately tell the overall energy of the bath. One therefore expects the interaction strength not to vary with the energy of the bath states that it couples. 

Because the bath states are necessarily more sparse in the lower part of the spectrum, and given that the interaction rate $g$ is limited, the lower states will not be as well mixed by the interaction, and the deviations from the thermal state will be larger. While unjustified, it is therefore tempting to increase the magnitude of the elements of $Y_{\mbox{\scriptsize bath}}$ so as to increase the mixing of the lower energy levels. This is what we do in our numerical simulation. In particular, if the element $Y_{ij}$ couples the bath energy levels $E_i$ and $E_j$, then we chose 
\begin{equation}
  Y_{ij} = g r_{ij} \left[1 + f\sqrt{(E_j-E_{j-1})(E_i-E_{i-1})} \right] , 
\end{equation}
where $r_{ij}$ is a Gaussian random number with mean zero and unit variance, $E_{k-1}$ is the energy level less than and adjacent to level $E_k$ (for all $k$), and we set $f = 100$.  

\begin{figure*}[t] 
\leavevmode\includegraphics[width=0.9\hsize]{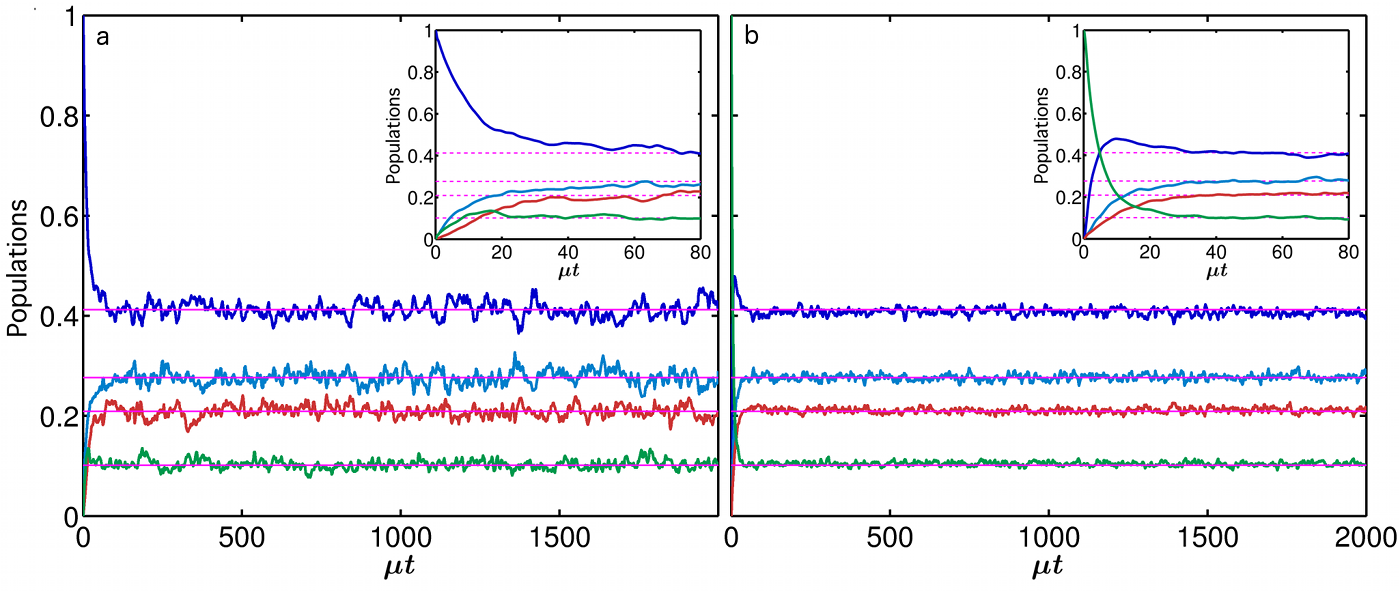} 
\caption{(Color online) The evolution of the populations of the energy-eigenstates of a nonlinear 4-level system coupled to a bath with 5,000 states, for two initial states of the system. The horizontal lines give the populations for the Boltzmann thermal distribution at the relevant temperature. In plot (a) the initial state is the one with lowest energy, and in (b) it is the state with the highest energy. The insets are expanded versions of the plots for early times, showing the initial relaxation to the thermal state. } 
\label{fig2} 
\end{figure*} 

6) The system: now that we have determined the structure of the bath, it is time to couple it to a small system. To limit the numerical overhead, we use a system with just four states. Naturally, the bath is required to thermalize any system, including any system interaction operator $X_{\mbox{\scriptsize sys}}$, with the only condition that $X_{\mbox{\scriptsize sys}}$ be sufficiently non-commuting with $H_{\mbox{\scriptsize sys}}$. Essentially any matrix will do for $X_{\mbox{\scriptsize sys}}$ so long as it provides enough transitions between the system states: it must be possible to go from any system state to any other system state by using a sequence of these transitions. We therefore make an arbitrary choice for the system energy levels and interaction operator $X_{\mbox{\scriptsize sys}}$. By ``arbitrary'' we mean that there is no special  relationship (symmetries) between the various energy gaps of the system, and between the elements of $X_{\mbox{\scriptsize sys}}$. If the bath correctly thermalizes such a system, then we can be confident that it will thermalize any system. We choose the energy levels of the system to be $\hbar \mu [0.5, 1.5, 2.2, 4]$. Denoting the matrix elements of $X_{\mbox{\scriptsize sys}}$ by $x_{ij}$, we choose $x_{12} = -0.7 $, $x_{13} = 0.3$, $x_{14} = -0.9$, $x_{23} = -1.2$, $x_{24} = -0.4 = -x_{34}$. The diagonal elements of $X_{\mbox{\scriptsize sys}}$ are set to zero, since there is no sense in unnecessarily perturbing the system.

\section{Eigenstate Thermalization}
\label{eigtherm}

We now explain why it is that the above model of a system-bath interaction can be expected not only to thermalize the system to the Boltzmann distribution, but to do so via the mechanism of eigenstate thermalization. Note that since many-body systems have been shown to thermalize via this mechanism~\cite{Rigol08}, if our model does so then it is correctly reproducing this behavior. Let us denote the energy eigenstates of the system by $|\varepsilon_k\rangle$, where $k = 1,\ldots, N_{\mbox{\scriptsize sys}}$, and those of the bath by $|E_j\rangle$, with $j = 1,\ldots, N_{\mbox{\scriptsize bath}}$. The energy eigenstates of the universe, before the interaction Hamiltonian is turned on, are then given by 
\begin{equation}
  | \mathcal{E}_{kj} \rangle = |\varepsilon_k\rangle |E_j\rangle , 
\end{equation}
where the total energy of the universe for each state is $\mathcal{E}_{kj} = \varepsilon_k + E_j$. As in the previous section we will refer to these states as the ``basis states'' of the universe. 

First we note that the interaction Hamiltonian, since it has elements of magnitude $\sim \hbar g$, will couple together (mix) only those basis states $| \mathcal{E}_{kj} \rangle$ whose energies are within approximately $\hbar g$ of each other. Consider now all the basis states within an energy band of width $2\hbar g$, centered at the energy $E_{\mbox{\scriptsize tot}}$. Each system state with energy $\varepsilon_k$ will appear in this set of basis states when it is paired only with bath states that have energies between  $E_{\mbox{\scriptsize tot}} - \varepsilon_k - \hbar g$ and $E_{\mbox{\scriptsize tot}} - \varepsilon_k  + \hbar g$. System states with lower energies are therefore paired with bath states that have higher energies. Since the number of bath states per unit energy increases exponentially, this means that there will be many more states in this set that contain the lowest system energy state, than higher system energy states. Let us denote the number of basis states in the band $[E_{\mbox{\scriptsize tot}}-\hbar g,E_{\mbox{\scriptsize tot}}+\hbar g]$ that contain system state $k$ as $N_k$. Then these numbers $N_k$ decrease exponentially with the system energy $\varepsilon_k$, exactly as the Boltzmann ratios:  
\begin{equation}
  \frac{N_m}{N_k} =  \exp[-\beta (\varepsilon_m - \varepsilon_k)] . 
\end{equation}
If the total number of basis states in the band is $N_{\mbox{\scriptsize band}}$, then the Boltzmann probability distribution is 
\begin{equation}
  P_k =  \exp[-\beta \varepsilon_k]/N_{\mbox{\scriptsize band}} . 
\end{equation}

Next we consider the eigenstates of the universe when the interaction is turned on. Let us denote these states by $|\tilde{E}\rangle$. The state $|\tilde{E}\rangle  $ is a superposition of the basis states $| \mathcal{E}_{kj} \rangle$ that have energies in the band $[E-\hbar g , E + \hbar g]$. The crucial point is that because the interaction Hamiltonian is random, one expects the state $|\tilde{E}\rangle$ to be a random superposition of all the basis states in the band. Because there are a large number of states in the band, the law of large numbers now tells us that the total contribution of the states that contain the system state with energy $\varepsilon_k$ will be approximately $N_k/N_{\mbox{\scriptsize band}}$. This is, of course, precisely the Boltzmann weighting, $P_k$. The larger $N_{\mbox{\scriptsize band}}$, then the more closely the contribution of the system state $|\varepsilon_k\rangle$ will be to $P_k$. If we now take the state $|\tilde{E}\rangle$, and trace out the bath, the contributions of the system states become the probabilities of the system states in the resulting mixture. That is 
\begin{equation}
  \mbox{Tr}_{\mbox{\scriptsize bath}} \left[ |\tilde{E}\rangle  \langle \tilde{E} | \right]  \approx  \sum_{k} \left( \frac{\exp(-\beta \varepsilon_k)}{\sum_n \exp(-\beta \varepsilon_n)} \right) |\varepsilon_k \rangle \langle \varepsilon_k |, 
\end{equation}
which is the thermal steady-state for the system. Thus we expect every energy eigenstate of the universe to give the Boltzmann state for the system, and this is eigenstate-thermalization. 

The above analysis also tells us that we can obtain a steady-state for the system that has a different  distribution over the energy states, by choosing the bath energy states to have a density profile equal to that new distribution (but reflected in energy). The density profile of the bath energy levels is copied onto the system steady-state, just as is predicted by the fundamental assumption of statistical mechanics. But note that here the origin is not the assumption that all states of in a given energy band are equally likely, but the assumption that the universe eigenstates are effectively random (within small energy bands), or equivalently that they are typical states within such bands~\cite{Popescu06, Goldstein06}.

\section{Numerical Simulations} 
\label{num}

To demonstrate thermalization we must evolve the system for an arbitrarily long time. Obtaining an essentially exact evolution for long times can be achieved by performing a full diagonalization of the Hamiltonian for the universe, $H_{\mbox{\scriptsize unv}}$. Since the Hamiltonian is a (real) $20,000$ dimensional matrix, this diagonalization does require a very large RAM. Nevertheless, with currently available computing resources, and absolute addressing, this is now quite feasible. In fact, we have already diagonalized real Hamiltonians that are twice this size, and even larger problems are clearly feasible.  

In Fig.~\ref{fig2} we present the results of the simulation, for two initial states of the system, being respectively the lowest and highest energy levels. Both initial states relax as desired to the thermal Boltzmann distribution and remain there, albeit with small fluctuations. Interestingly, when the system starts in its ground state, the residual fluctuations are larger than when it starts in its highest energy state. We will return to this phenomena below, which is due to the finite size of the bath. 

\begin{figure*}[t] 
\leavevmode\includegraphics[width=0.9\hsize]{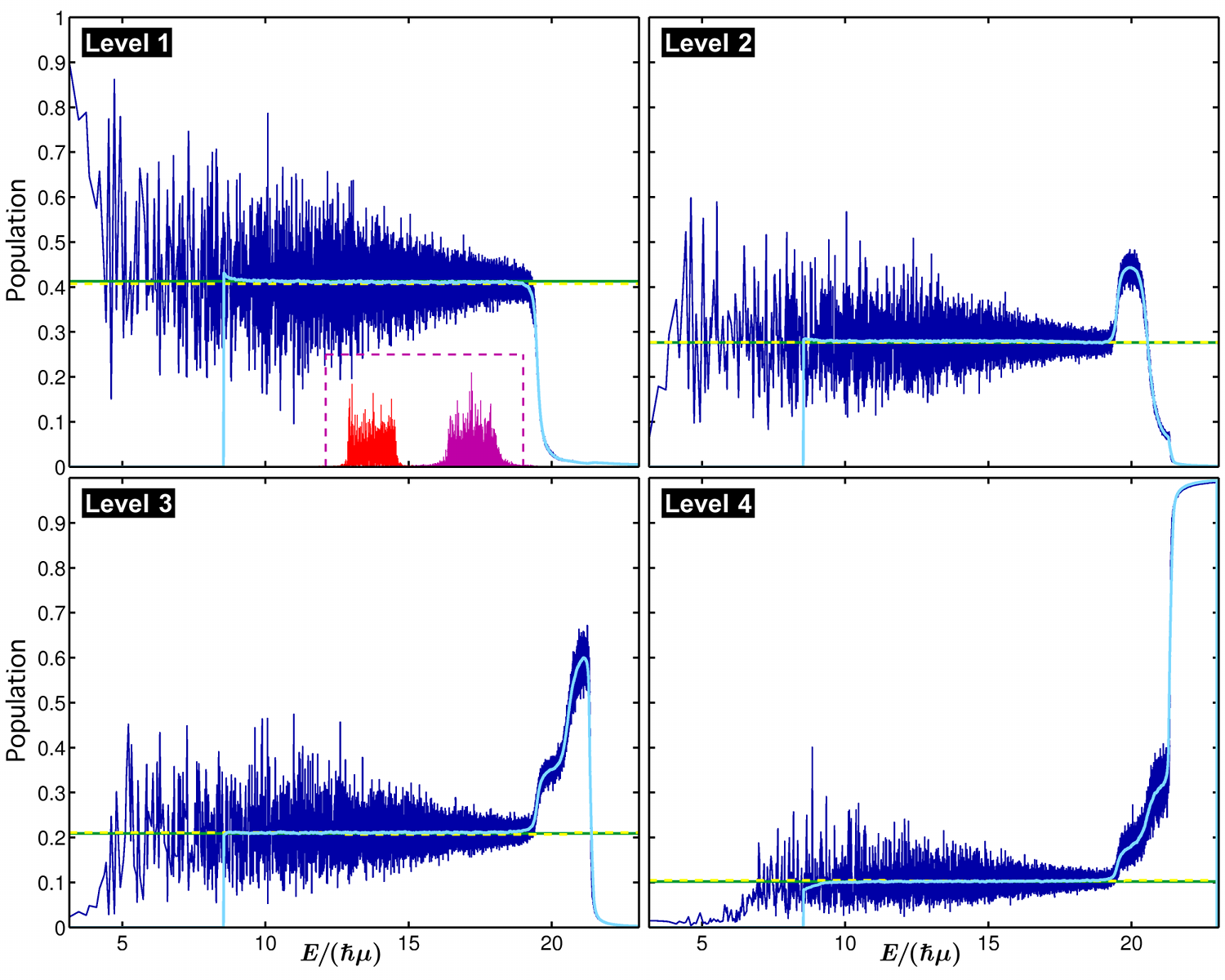} 
\caption{(Color online) Each of the four plots shows the populations of one of the system energy levels (of which there are four). In each plot: The dashed horizontal line (yellow) is the desired thermal (Boltzmann) value. The light-grey solid horizontal line (green) is the actual steady-state population when the initial state is state 4. The dark noisy line (dark blue) is the population given by each of the energy eigenstates of the universe, as a function of their energy. The solid light grey curve (cyan) is the moving average of the noisy line over a window of 200 adjacent eigenstates. Thermalization happens in the region where the  light grey curve (cyan), light grey horizontal line (yellow) and horizontal dashed line (green) coincide. In the dashed box in the plot for level 1 we display the distribution of the initial state of the universe, over its eigenstates, when the system starts in state 1 (left, red) and 4 (right, mauve).} 
\label{fig3} 
\end{figure*} 

We now turn to the question of eigenstate thermalization~\cite{Srednicki94, Rigol08}. If eigenstate thermalization occurs, then for each eigenstate of the universe, the reduced state of the system (that is, traced over the bath) will be the thermal Boltzmann state. We will refer to the population for a system energy eigenstate that results from the universe being in a single energy eigenstate, as the ``eigenstate-value'' for that population. These ``eigenstate-values'' are shown in the four plots in Fig.~\ref{fig3}. In each of the four plots, the horizontal solid line gives the Boltzmann population for the respective state. The dark (noisy) line gives the eigenstate-values for the population as a function of the energy of the eigenstates. We see that the eigenstate-values are not in fact equal to the thermal value, since they fluctuate significantly. However, we expect precisely such fluctuations if the number of states within each energy band (the number coupled together by the interaction) is not sufficiently large (see Section~\ref{eigtherm}). To make $N_{\mbox{\scriptsize band}}$ large, the energy separation between adjacent bath states must be much less than $\hbar g$. The difficulty is that the exponential form of the density profile, together with the need for the bath to span a sufficient energy range, forces the energy spacings between the lower energy bath states to be relatively large. We would therefore expect eigenstate-thermalization to be true for the highest energy states, and the deviations from the Boltzmann populations to increase as the energy decreases. This is precisely what we see in Fig.~\ref{fig3}. Note that increasing the interaction strength $g$ will increase $N_{\mbox{\scriptsize band}}$, and thus reduce the fluctuations of the universe eigenstates. But recall that $g$ is also limited by the requirement $g \ll \mu$. 

A quantitative measure of the degree to which eigenstate thermalization is realized in a particular system is given in Eq.(6) of~\cite{olshanii2012_0582}. This measure, which we will call $\xi$, is the ratio of the variance across the eigenstates of the quantum mean of an observable $A$ for each eigenstate to the mean across the eigenstates of the quantum variance of $A$ for each eigenstate. 
Eigenstate thermalization is achieved when $\xi \ll 1$. If we denote the eigenstates by $|E_n\rangle$, and the initial probability that the universe is in state $|E_n\rangle$ by $p_n$, then 
\begin{equation}
  \xi = \frac{\sum_n p_n \langle E_n|A|E_n\rangle ^2 - \left[ \sum_n p_n \langle E_n|A|E_n\rangle \right]^2}{ \sum_n p_n \left[ \langle E_n|A^2|E_n\rangle - \langle E_n|A|E_n\rangle^2 \right] }
  \label{eqxi}
\end{equation}
Calculating $\xi$ for the population of level 1 (that is, choosing $A$ to be the projector onto level 1), using the data displayed in Fig.~\ref{fig3} and averaging over the eigenstates with $E/(\hbar\mu) \in [10, 15]$, we find that $\xi = 0.02$. This shows us that the universe has achieved eigenstate thermalization to a significant degree. 

\begin{figure}[t] 
\leavevmode\includegraphics[width=1\hsize]{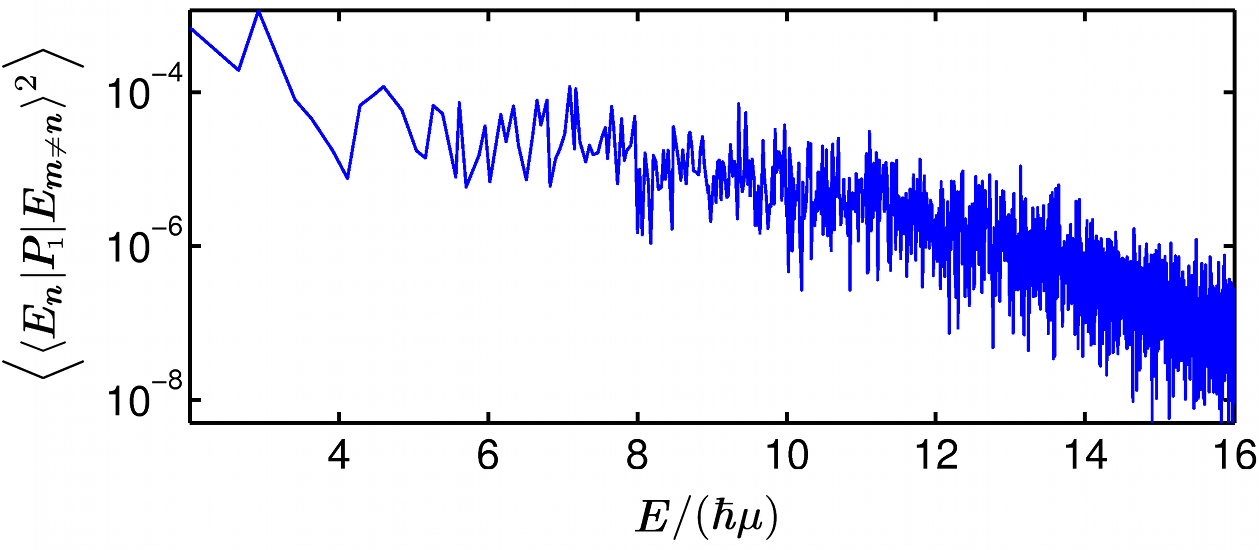} 
\caption{(Color online) Here we plot an estimate of the mean square of the off-diagonal elements of the projector onto the ground state, $P_1 = |1\rangle \langle 1|$. This estimate is calculated, for each value of $n$, by averaging the square of the off-diagonal elements for $m = n+1$ to $m = n+10$.} 
\label{fig4} 
\end{figure}

If we decrease $\beta$, and thus increase the temperature, the density profile of the bath energy states becomes more even, allowing the lowest bath energy levels to be more closely spaced. This increases the mixing that can be achieved across all the states of the bath, and eigenstate thermalization is approached more closely. Thus the higher the temperature, the easier it is to realize thermalization, in that smaller baths will suffice. For any temperature, as the size of the bath is increased, the separation of the energy levels decreases, increasing the mixing and reducing the fluctuations in the populations from one eigenstate to the next. In the limit of a macroscopic bath, the fluctuations of the populations of the eigenstates tends to zero, and true eigenstate thermalization will be achieved.  

We also obtain insight into why the average of the eigenstate-values gives the correct thermal value, independent of the extent to which eigenstate-thermalization is realized. The key to this behavior can be seen by turning of the interaction (setting $g=0$). The eigenstates of the universe are now merely the tensor-product states of the system eigenstates and the bath eigenstates. But the average of the eigenstate-values of these tensor-product states is \textit{still} equal to the thermal value, precisely because the bath has an exponential density of states. That is, even before we turn the interaction on, the eigenstate-values are already what they need to be. To realize thermalization all the interaction has to do is to mix the tensor-product states sufficiently so that each universe eigenstate is a superposition over a sufficiently large number of the tensor-product states, while preserving the average of the eigenstate-values. 

We return now to the question of why the fluctuations of the populations are significantly smaller when the system starts in its highest energy state, as opposed to its ground state, as evident in Fig.~\ref{fig2}. The answer lies in the fact that we chose the same energy window for the bath in both cases. As a result, in the former case, the state of the universe covers a higher energy window, a window over which the density-of-states of the bath is higher. The higher the density of states, the more we expect the averaging process (the mixing of the states due to the interaction) to reduce the fluctuations. In particular, if we use a similar method to that in~\cite{Srednicki94},
we can show that the variance of the temporal fluctuations of the quantum mean of a Hermitian operator $A$ is approximately  
\begin{equation}
   V_{\mbox{\scriptsize ss}}(A) = \sum_{n,m\not= n} p_n p_m \langle E_n | A | E_m \rangle^2 , 
\end{equation}
where we are using the same definitions as in Eq.(\ref{eqxi}) above. The temporal fluctuations of the system populations will therefore decrease as the off-diagonal elements of the projectors onto the system states decrease. In Fig.~\ref{fig4} we display an average of the mean-square of these off-diagonal elements, for the projector onto the ground state, as a function of energy of the bath. This shows, as expected, that these elements decrease as the density of bath states increases. 

\section{Entropy production and entanglement}
\label{ent}
We now examine the increase in thermodynamic entropy, both of the bath and the system, associated with the (effectively) irreversible evolution. In fact, the relationship between the von Neumann entropy, $S_{\ms{vN}}$, and the thermodynamic entropy, $S_{\ms{th}}$, is not the same for the system and the bath. Nor does entanglement play the same role for each. This is because the micro states of the system are accessible (by definition), whereas those of the bath are not, and this inaccessibility is part of the definition of the latter's thermodynamic entropy. We first consider the thermodynamic entropy of the system. The system starts in a non-equilibrium state, and we must therefore be careful to specify the context in which we define its thermodynamic entropy --- different contexts may motivate different definitions. For a standard thermodynamic system in an equilibrium state, and in contact with a thermal bath at temperature $T$, the maximum work that can be extracted by manipulating the system is given by the free energy, $F = E - TS_{\ms{th}}$, where $E$ is the internal energy and $T$ is the temperature of the bath. It has been shown that for a quantum system in contact with a thermal bath at temperature $T$, in which all unitary operations on the system are available to the controller (and thus all micro states of the system are accessible), the maximum work that can be extracted is given by $\mathcal{F} = \langle E \rangle - T S_{\ms{vN}}(\rho)$, where $\rho$ is the density matrix of the system~\cite{Hasegawa10, Takara10, Esposito11}. This is true for all states, equilibrium or otherwise. Thus, in the context of work extraction, the von Neumann entropy of a mesoscopic quantum system can be identified with its thermodynamic entropy. 

If the bath starts in a pure state, which is well-motivated from a fundamental point of view, then since the evolution is unitary, any increase in the von Neuman entropy of the system can only be generated by entanglement between the system and bath. Further, if the joint state of two systems is pure, then a good measure of the entanglement between the two is the von Nuemann entropy of either~\cite{Bennett96}. Since the steady-state of the system is the Boltzman state, the total entropy produced in the system is simply the difference between the initial von Neumann entropy of the system and the von Neumann entropy of the Boltzmann state. This increase in entropy is entirely entropy \textit{production} because no work is extracted in the process, so that the free energy lost cannot be regained. 

The thermodynamic entropy of the bath is $k_{\ms{B}} \ln(N_{\ms{acc}})$, where $N_{\ms{acc}}$ is the number of accessible micro states for the given macro state. To put an absolute value on this entropy we need to fix the width of an energy window that we consider to be the region of accessible energy. This ``course grains'' the micro states into macro states. One way to define this width is as the average size of the elements of the interaction Hamiltonian that couples the system and bath. This energy scale gives the energy width over which the bath eigenstates are coupled together, and thus determines the number of bath eigenstates that are explored during the evolution. But we will not concern ourselves here with the absolute value of the entropy, merely the increase in the entropy of the bath during the thermalization of the system. 

We note that under the above definition, the thermodynamic entropy of the bath has nothing to do with the von Neumann entropy. We make a connection with the von Neumann entropy only if we consider the state-of-knowledge of a macroscopic observer that is ignorant of the micro state up to the course-graining specified above. In this case the thermodynamic entropy is given by ($k_{\ms{B}}$ times) the von Neumann entropy of the macroscopic observer's state of knowledge. But there is no need to introduce such an observer. Since the density of bath energy levels increases by the factor $e^{\Delta E /k_{\ms{B}} T}$ with an increase in energy $\Delta E$, given a fixed width for the energy window, an increase of energy $\Delta E$ increases the bath's thermodynamic entropy by $\Delta S_{\ms{th}} = \Delta E /T$, in accordance with standard thermodynamics. Now let us see how this plays out in more detail, given the final (equilibrium) state of the system. This final state is a mixture over the energy eigenstates of the system. Let us say, for simplicity, that the initial state of the system is a single energy eigenstate. For each final eigenstate of the system, $|\varepsilon_i\rangle$, the energy of the bath has been changed by the negative of the change to the energy of the system, which we denote by $\Delta \varepsilon_i$. The change in the entropy of the bath for each final system eigenstate is then $-\Delta \varepsilon_i/T$. The total change in the entropy of the bath is the average of these changes over the final probabilities of the system eigenstates, $-\langle \Delta\varepsilon\rangle/T$. We note that if we write the final state of the bath as the state-of-knowledge of an observer who is ignorant of the micro states under the course-graining, but knows the final density matrix of the system, then the total von Neumann entropy change of the state of the universe is precisely the sum of the von Neumann entropy change of the system, and that given above for the thermodynamic entropy change of the bath (divided by $k$), due to the well-known course-graining property of the entropy~\cite{Jacobs14}. Thus the definitions of entropy of the system and bath are consistent.  

To summarize, the thermodynamic entropy production for the system is 
\begin{equation}
 \Delta S_{\ms{th}}^{\ms{sys}} =  k_{\ms{B}} \Delta S_{\ms{vN}} \;\;\;\; \mbox{(due entirely to entanglement)} , 
\end{equation}
and that for the bath is 
\begin{equation}
 \Delta S_{\ms{th}}^{\ms{bath}} =  -\frac{\langle \Delta \varepsilon\rangle}{T}  \;\;\;\;  \mbox{(unrelated to entanglement)} ,  
\end{equation}
where $\langle \Delta \varepsilon \rangle$ and $\Delta S_{\ms{vN}}$ are, respectively, the average change in the energy of the system, and the change in the von Neumann entropy of the system. The sum of these two entropy changes is always non-negative, which follows from the results in~\cite{Hasegawa10, Esposito11}. 

\section{Reproducing the Redfield evolution, and the question of universality}
\label{relax}

Even when we use a bath containing 5,000 states, there are still significant fluctuations in the relaxation dynamics. We can reduce these fluctuations by averaging the evolution over many randomly chosen initial states of the bath. The goal of reducing the fluctuations is to obtain a better approximation to the dynamics induced by a bath with an infinite number of states. After averaging away the fluctuations, we can compare the relaxation dynamics for different bath sizes, and determine how many bath states are sufficient to reproduce the behavior of a macroscopic bath. We can also compare this thermal relaxation to that of the Markovian Redfield master equation~\cite{Redfield57, Redfield65}. This Redfield equation is derived using a bath consisting of a continuum of harmonic oscillators, quite different from our random bath. However, the Redfield equation is a perturbative master equation, valid for weak damping, and as such is a set of rate equations, where the rates are determined by Fermi's golden rule. Thus we expect that for weak damping our bath should agree with the Redfield equation, since Fermi's golden rule is a result of the near-continuum of the energy levels of the bath, and our system-bath coupling is essentially generic as far as the bath is concerned. 

The Redfield master equation gives the following rate equations for the populations of the system~\cite{Redfield57, Breuer07}: 
\begin{equation}
    P_{j} = - \left( \sum_i  \gamma_{i\leftarrow j} \right) P_{j} + \sum_i \gamma_{j\leftarrow i} P_{i}
\end{equation}
where $\gamma_{j \leftarrow i}$ is the transition rate from level $i$ to level $j$. We note that the Markovian Redfield equation is a result of the rotating-wave approximation (secular approxmation) valid for weak coupling. Weak coupling is defined by $\gamma_{j\leftarrow i} \ll |\varepsilon_{i} - \varepsilon_j|$, for all $i,j$, where as above $\varepsilon_m$ denotes energy level $m$ of the system.  

Applying Fermi's golden rule to our model, the transition rates resulting from our system/bath coupling should be 
\begin{equation}
    \gamma_{j\leftarrow i} = 2\pi |x_{ij}|^2 \langle |Y|^2_{j \leftarrow i} \mathcal{\rho}_{j \leftarrow i} \rangle ,  
\end{equation}
where $|Y|^2_{j \leftarrow i}$ is the square of an element of the bath interaction operator that couples an initial state of the bath to the corresponding final state, for the transition $j\leftarrow i$, and $\rho_{j \leftarrow i}$ is the density of the final bath states with respect to energy for this bath transition. The values of $|Y|^2_{j \leftarrow i}$ and $\rho_{j \leftarrow i}$ depend on the initial state of the bath, so we must average these quantities over the initially populated states of the bath.  

\begin{figure}[t] 
\leavevmode\includegraphics[width=1\hsize]{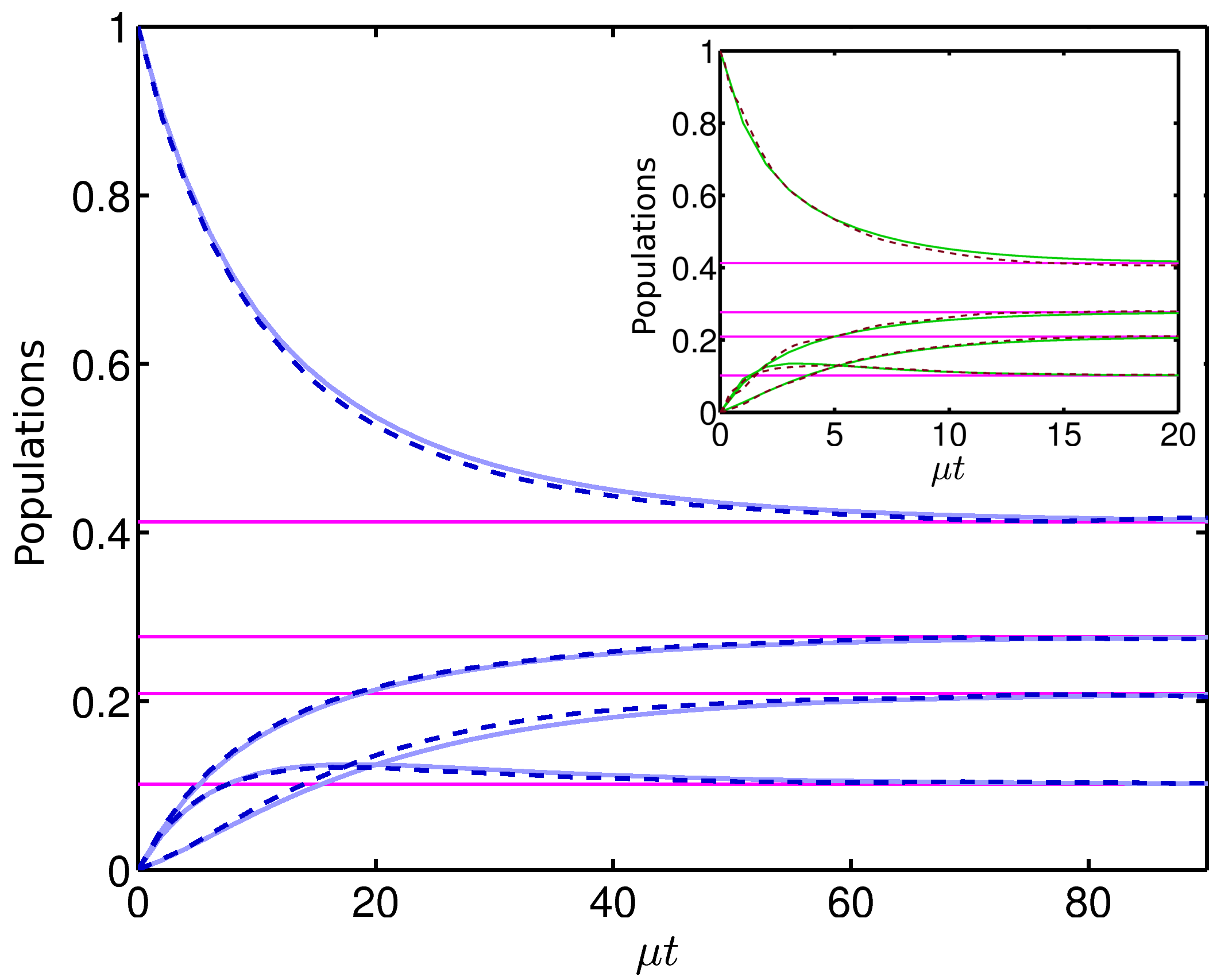} 
\caption{(Color online) Here we plot the dynamics of the thermal relaxation of the system, averaged over 200 initial states of the bath, randomly chosen within the same energy window (dashed lines). This averages out the fluctuations due to the finite size of the bath, and we obtain a good approximation to the evolution for the macroscopic bath. We also plot the thermal relaxation given by the equivalent Markovian Redfield master equation (solid line). The main plot is for a bath with 5,000 states, and the inset gives the result for a bath of 2,500 states.} 
\label{fig5} 
\end{figure} 

In Fig.~\ref{fig5} we show the evolution for a bath of 5000 states, averaged over 200 initial states of the bath, randomly chosen within the fixed energy window used above, and the inset is for a bath of 2500 states. Both cases agree quite well with the Redfield rate equations, although fluctuations are more pronounced for the smaller bath. Thus our bath is able to give a good approximation to the dynamics induced by an infinite (macroscopic) Markovian bath with the same structure. 

Since all baths have a near-continuum of states, one might be temped to assume that the dynamics of the damping of open systems coupled to thermal baths is universal, but this is not the case. Simulations of a small system coupled to a single spin that forms one end of a spin chain (a one-dimensional spin lattice) show that the relaxation in this case is not merely the simple exponential decay generated by rate equations, but depends on the details of the coupling between the system and the spin on the end of the chain, as well as the coupling between the spins. This behavior is quite reasonable: in finite time the evolution of the system can only be affected by a finite number of the spins in the chain, since only nearest-neighbors are coupled. Because of this only a small number of spins contribute to the evolution of the system during the initial relaxation, and so the dynamics is determined by the local coupling Hamiltonians. Typicality determines the steady-state, but cannot determine the initial relaxation. 

We can conclude from the above discussion that the damping induced by many-body baths can only be universal if the system is coupled to a large number of the bodies. In this case, even if the system is weakly coupled to each body, the result can be either weak or strong damping. If in either case the  dynamics of the thermal relaxation turns out to be universal, then it will be useful to determine what models correctly reproduce this dynamics. Since we know that many-body systems that thermalize do have a density of states that is exponential in the energy, and have typical eigenstates, the random bath is a good candidate for such a model.  While the oscillator-bath model has become, by default, the gold-standard for describing the damping of both weakly and strongly damped systems (e.g. Brownian motion), there is little justification for this special status. To obtain Brownian motion using an oscillator-bath one must choose the coupling to have an ``Ohmic'' dependence on frequency, a rather arbitrary choice. The efficacy of any model of thermal relaxation will depend on the extent to which the dynamics of relaxation is universal, and the extent to which it reproduces this relaxation. These questions are interesting topics for future work.

\section{Summary}
\label{sumup}

We have shown that the thermal relaxation of a small system can be modeled, to good approximation, by exactly simulating the evolution of the system coupled to a bath containing a few thousand states. To do so, we chose the bath so as to have certain key quantities possessed by many-body systems, and we average the resulting time evolution for the system over a few hundred initial states of the bath. The latter procedure greatly reduces the fluctuations in the evolution of the system due to the finite size of the bath. 

We suggest that the model we have presented is a good candidate for simulating the evolution of small systems strongly coupled to real many-body systems, because it is derived using the assumption of typicality. Of course, any generic model of thermalization will reproduce the damping induced by many-body systems only for classes of couplings for which this damping is universal. What these classes may be is as yet an open question.

\section*{Acknowledgments}

The authors are indebted to Prof.\ Daniel Steck at the University of Oregon and the Oregon Center for Optics; this work would not have been possible without access to the large memory nodes of his parallel cluster, funded by the National Science Foundation under Project No.\ PHY-0547926. In the latter stages of this work we used the supercomputing facilities managed by the Research Computing Group at the University of Massachusetts Boston. During this work KJ was partially supported by the National Science Foundation under Project Nos.\ PHY-0902906, PHY-1005571, and PHY-1212413, and the ARO MURI grant W911NF-11-1-0268. LS was supported by the NSF under Project No. PHY-0902906, and MO and VD were supported by the  Office of Naval Research grant no.\ N00014-12-1-0400 and the National Science Foundation grant no.\ PHY-1019197. 


\end{document}